\documentclass[twocolumn,showpacs,amsmath,amssymb,pra]{revtex4}
\usepackage{amsxtra}
\usepackage{amssymb}
\usepackage{amsmath}
\usepackage{graphicx}

\begin{document}
\title{Comment on ``Observation of anticorrelation in incoherent thermal light fields''}
\date{January 6, 2012}
\author{Jeffrey H. Shapiro}
\affiliation{Research Laboratory of Electronics, Massachusetts Institute of Technology, Cambridge, Massachusetts 02139, USA}
\author{Eric Lantz}
\affiliation{D\'{e}partement d'Optique P.M. Duffieux, Institut
FEMTO-ST, UMR CNRS 6174,\\ Universit\'{e} de Franche-Comt\'{e}, 16
route de Gray, F-25030 Besan\c{c}on Cedex, France}
\begin{abstract}
Recently, Chen \em et al\/\rm.\ [Phys. Rev. A {\bf 84,} 033835 (2011)] reported observation of anticorrelated photon coincidences in a Mach-Zehnder interferometer whose input light came from a mode-locked Ti:sapphire laser that had been rendered spatially incoherent by passage through a rotating ground-glass diffuser.  They provided a quantum-mechanical explanation of their results, which ascribes the anticorrelation to two-photon interference.  They also developed a classical-light treatment of the experiment, and showed that it was incapable of explaining the anticorrelation behavior.  Here we show that semiclassical photodetection theory---i.e., classical electromagnetic fields plus photodetector shot noise---does indeed explain the anticorrelation found by Chen \em et al\/\rm.\  The key to our analysis is proper accounting for the disparate time scales associated with the laser's pulse duration, the speckle-correlation time, the interferometer's differential delay, and the duration of the photon-coincidence gate.  Our result is consistent with the long-accepted dictum that laser light which has undergone linear-optics transformations is classical-state light, so that the quantum and semiclassical theories of photodetection yield quantitatively identical results for its measurement statistics.  The interpretation provided by Chen \em et al\/\rm.\ for their observations implicitly contradicts that dictum.

\end{abstract}

\maketitle

The recent paper by Chen \em et al\/\rm.\ \cite{Chen} reports the following experiment.  A continuous-wave mode-locked Ti:sapphire laser operating at $\lambda = 780$\,nm wavelength with 78\,MHz pulse-repetition frequency, and a $\tau_p\sim$150\,fs pulse duration illuminated an interference filter, to somewhat increase the pulse duration, followed by a rotating ground-glass diffuser, to render the light spatially incoherent.  The diameter $D=4.5$\,mm output beam from the diffuser was divided by a 50-50 beam splitter, with
the resulting beams propagating $d\approx 200$\,mm (from the diffuser) to collection planes, each of which contained the tip of a single-mode optical fiber.  These fibers routed the light they collected to another 50-50 beam splitter whose outputs illuminated single-photon detectors.  By sufficiently offsetting, in their respective planes, the transverse coordinates of the fiber tips that collected light from the diffuser, Chen \em et al\/\rm.\ ensured that there was no first-order interference in the Mach-Zehnder interferometer formed by the two 50-50 beam splitters and the intervening fibers.  They measured photon coincidences between the two detectors, with a $T\sim 1$\,ns coincidence gate, as one of the collection fibers was moved longitudinally to create a $-$2\,ps $\le \delta t\le 2$\,ps differential delay.  What they observed was a pronounced dip---a photon anticorrelation---in the coincidence rate, despite the absence of any delay-dependence in the singles rates.  See \cite{Chen} Fig.~3 for a diagram of the Chen \em et al\/\rm.\ experiment, and \cite{Chen} Fig.~4 for their observations of anticorrelation.

Chen \em et al\/\rm.\ provided a quantum-mechanical explanation for the anticorrelation seen in their experiment, which shows that it is due to two-photon interference.  Because light is quantum mechanical, and photodetection is a quantum measurement, there must be a quantum explanation for the results in \cite{Chen}.  But the authors of \cite{Chen} do more than provide a quantum explanation for their observations.  They presents a classical-field analysis that, they claim, proves that \em only \/\rm\ a quantum treatment can account for the anticorrelation they found.
Were these authors correct, their work would present a very significant conundrum for quantum optics.  Laser light, except for any excess noise it may carry, is coherent-state light.  Passage through a ground-glass diffuser, free-space propagation, beam splitting, and fiber propagation are all linear optical effects, with the first best modeled as a random process while the rest can be taken to be deterministic.  Taken together, the preceding two sentences imply that the joint quantum state of the fields illuminating the two detectors in Fig.~3 of Chen \em et al\/\rm.\ is \em classical\/\rm, viz., it is a random mixture of coherent states.  It has long been known that the quantum and semiclassical \cite{footnote1} theories of photodetection yield quantitatively \em identical\/\rm\ predictions for classical-state illumination, see \cite{Shapiro} for a detailed review of this topic.

So, in view of the preceding discussion, we can say that one of three things must be true:  (1) despite what is argued in \cite{Chen}, there \em is\/\rm\ a classical explanation for the anticorrelation reported therein;  or (2) laser light that has undergone linear transformation is \em not\/\rm\ in a coherent state or a random mixture of coherent states; or (3) the quantum and semiclassical theories of photodetection \em can\/\rm\ make different quantitative predictions for the measurement statistics of classical-state illumination.  To assert the truth of items (2) and/or (3), as Chen \em et al\/\rm.\ implicitly do, would constitute a major upheaval in quantum optics.  We shall show that item (1) holds.  The key to doing so is proper accounting for the disparate time scales associated with the laser's pulse duration, the speckle-correlation time, the interferometer's differential delay, and the duration of the photon-coincidence gate.

With the interference filter in place, the duration of the laser pulse that illuminated the ground-glass diffuser in Fig.~3 of \cite{Chen} was increased to either $\tau_p \sim 345\,$fs or $\tau_p \sim 541$\,fs, depending on which of two interference filters was employed.  The linear velocity of the rotating ground-glass where it was illuminated was $\sim$0.8\,m/s \cite{Chen}, so that for either interference filter it is fair to assume that the ground glass was completely stationary while a single laser pulse propagated through it.  In other words, the speckle correlation time greatly exceeded $\tau_p$.  The differential delay over which Chen \em et al\/\rm.\ traced out coincidence rates was $|\delta t|\le 2$\,ps.  Thus the duration of the photon-coincidence gate in Fig.~3 of \cite{Chen} obeyed $T \gg |\tau_p \pm \delta t|$.

Chen \em et al\/\rm.\ used single-mode fibers to collect spatial samples of the two light beams that had propagated $d\approx 200$\,mm from the diffuser and been separated by the initial 50-50 beam splitter in their Fig.~3.  Coherence theory \cite{MandelWolf} shows that the fields at that distance from the diffuser have $\ell_c\sim\lambda d/D \approx 35\,\mu$m transverse coherence lengths \cite{footnote2}. The data in Fig.~4 of \cite{Chen} was collected with more than $40 \ell_c$ transverse separation, in their respective collection planes, between the tips of the single-mode fibers, whose core diameters we shall assume to be much smaller than $\ell_c$.  Hence the field injected in each fiber comes from a unique coherence cell, in time as well in space.  This ensures that every fiber-collected femtosecond pulse is coherent, although with a random phase and amplitude.  Moreover, the pulses in each fiber arise from different coherence cells, and so their random behaviors are statistically independent.  Nevertheless, it is incorrect to assert (cf.\ Sec.~4 of \cite{Chen}) that the light beams emerging from the two fibers do not interfere. Rather, they produce fringes that are random between pulses separated by more than the decorrelation time of the pseudothermal source.  More importantly, energy conservation implies there will be anticorrelation at output ports~1 and 2 in the Chen \em et al\/\rm. experiment, viz.,  a bright fringe in port~1 is always accompanied by a dark fringe in port~2.  As noted in \cite{Chen}, this anticorrelation would not depend on the interferometer's differential delay for continuous-wave (statistically stationary) pseudothermal light.  Chen \em et al\/\rm., however, used femtosecond pulses, for which  the anticorrelation disappears when the pulses do not overlap in time at the second 50-50 beam splitter, and this loss of anticorrelation occurs even though the necessary differential delay is much shorter than the photodetectors' nanosecond coincidence window.

The argument presented in the preceding paragraph constitutes a complete explanation of the Chen \em et al\/\rm. anticorrelation in terms of classical interference behavior.  We will now expand upon that classical-field explanation to provide a full quantitative treatment.  
We define $E_+(t)$ and $E_-(t)$ to be the $\sqrt{\mbox{photons/s}}$-units positive-frequency \em classical\/\rm\ fields entering the single-mode fibers from a single pulse occurring at time $t=0$ \cite{footnote3}.  Given that the speckle is frozen over a single laser pulse, and that the fibers have core diameters which are much smaller than $\ell_c$, it is fair to write these fields as follows:
\begin{equation}
E_{\pm}(t) = v_\pm f(t\pm\delta t/2)e^{-i\omega_0t},
\label{Epm}
\end{equation}
where $v_+$ and $v_-$ are independent, identically distributed, zero-mean, isotropic, complex-valued Gaussian random variables with common mean-squared strength
\begin{equation}
\langle |v_+|^2\rangle = \langle |v_-|^2\rangle = N,
\end{equation}
and
\begin{equation}
f(t) \equiv \frac{e^{-t^2/\tau_p^2}}{(\pi\tau_p^2/2)^{1/4}},
\label{pulse}
\end{equation}
is a transform-limited Gaussian pulse normalized to satisfy
\begin{equation}
\int\!dt\,|f(t)|^2 = 1.
\end{equation}
Physically, $v_+$ and $v_-$ are the constant-in-time speckle values for the given laser pulse, whose independence is guaranteed by the large transverse separation of the fibers in their respective collection planes.  Our $f(t)$ normalization then implies that $N\hbar \omega_0$, with $\omega_0 = 2\pi c/\lambda$, is the average energy entering each of the fibers from the given laser pulse.  Thus $N$ measures the average energy of these classical fields in photon units and, because the measurements reported in \cite{Chen} were made in the photon-counting regime, we will assume $N \ll 1$.  The fields that illuminate the photodetectors, which will denote $ E_1(t)$ and $ E_2(t)$, as was done in \cite{Chen}, are then given by
\begin{equation}
 E_1(t) \equiv \frac{E_+(t) + E_-(t)}{\sqrt{2}}
\label{EaDefn}
\end{equation}
and
\begin{equation}
 E_2(t) \equiv \frac{E_+(t) -E_-(t)}{\sqrt{2}}
\label{EbDefn}.
\end{equation}
Furthermore, because $N\ll 1$, we can say that the average singles rates (counts/gate) and coincidence rate (coincidences/gate) obey \cite{Shapiro}
\begin{equation}
S_K = \eta\int_{-T/2}^{T/2}\!dt\,\langle |E_K(t)|^2\rangle\quad\mbox{for $K= 1,2$,}
\end{equation}
and
\begin{equation}
C_{12} = \eta^2\int_{-T/2}^{T/2}\!dt\int_{-T/2}^{T/2}\!du\,\langle | E_1(t)|^2| E_2(u)|^2\rangle,
\end{equation}
where $\eta$ is the photodetectors' quantum efficiency.   All that remains is to evaluate these rates.

Using the statistical independence of $v_+$ and $v_-$ and their common mean-squared value, we immediately find that
\begin{eqnarray}
\lefteqn{S_1 = S_2 =}  \nonumber \\[.1in]
&&  \frac{\eta N}{2} \int_{-T/2}^{T/2}\!dt\,(|f(t+\delta t/2)|^2 + |f(t-\delta t/2)|^2)\\[.1in] &\approx& \eta N,
\end{eqnarray}
where the approximation follows from $|\tau_p \pm \delta t/2| \ll T$ and Eq.~(\ref{pulse}).
Similarly, for the coincidence rate,  the statistical independence of $v_+$ and $v_-$  leads to \cite{footnote4}
\begin{eqnarray}
C_{12} &=& \frac{\eta^2}{4}\int_{-T/2}^{T/2}\!dt\,\int_{-T/2}^{T/2}\!du\,[\langle |v_+|^4\rangle |f(t_+)|^2|f(u_+)|^2 \nonumber\\[.1in]
&+& \langle |v_+|^2\rangle \langle |v_-|^2\rangle |f(t_+)|^2|f(u_-)|^2 \nonumber \\[.1in]
&+&  \langle |v_+|^2\rangle \langle |v_-|^2\rangle |f(u_+)|^2|f(t_-)|^2 \nonumber\\[.1in]
&-&2\langle |v_+|^2\rangle \langle |v_-|^2\rangle{\rm Re}[f^*(t_+)f^*(u_-)f(t_-)f(u_+)] \nonumber \\[.1in]
&+& \langle |v_-|^4\rangle |f(t_-)|^2|f(u_-)|^2],
\label{C12dev}
\end{eqnarray}
where $t_\pm \equiv t \pm \delta t/2$ and $u_\pm \equiv u \pm \delta t/2$.  Now, using the Gaussian moment-factoring theorem \cite{WJ}, $|\tau_p\pm\delta t/2| \ll T$, and Eq.~(\ref{pulse}), we can reduce the preceding expression to
\begin{equation}
C_{12} \approx \frac{\eta^2N^2}{2}\left(3-\left|\int_{-T/2}^{T/2}\!d\tau\,f^*(\tau+\delta t/2)f(\tau-\delta t/2)\right|^2\right).
\end{equation}
 Using $|\tau_p\pm \delta t| \ll T$ and Eq.~(\ref{pulse}) then gives us our final result,
 \begin{equation}
 C_{12} \approx \frac{\eta^2N^2}{2}\left(3-e^{-\delta t^2/\tau_p^2}\right).
 \label{Cab}
 \end{equation}
 
 Equation~(\ref{C12dev}) can be obtained in a slightly different way to emphasize the presence of anticorrelated fringes at the output ports. The intensities at these ports can be obtained from Eqs.~(\ref{EaDefn}) and (\ref{EbDefn}) as:
 \begin{eqnarray}
 | E_1(t)|^2&=& \frac{1}{2}(|v_+|^2 |f(t_+)|^2+|v_-|^2 |f(t_-)|^2\nonumber \\[.1in]
 &&+2|v_+||v_-|\,{\rm Re}[f^*(t_+) f(t_-)e^{i\Delta\varphi}]\label{fringes1} \\[.1in]
 | E_2(u)|^2&=& \frac{1}{2}(|v_+|^2 |f(u_+)|^2+|v_-|^2 |f(u_-)|^2\nonumber \\[.1in]
 && -2|v_+||v_-|\,{\rm Re}[f^*(u_+) f(u_-)e^{i\Delta\varphi}]
\label{fringes2}
\end{eqnarray}
where $\Delta\varphi \equiv \varphi_--\varphi_+$ in terms of the phases, $\varphi_+$ and $\varphi_-$, associated with  $v_+$ and $v_-$.  Equation~(\ref{C12dev}) can be retrieved from Eqs.~(\ref{fringes1}) and (\ref{fringes2}) by noting that the amplitude and phase of $v_\pm$ are statistically independent, with $\varphi_\pm$ being uniformly distributed on $0\le \varphi_\pm \le 2\pi$, so that $\langle e^{i\varphi_\pm}\rangle = \langle e^{i2\varphi_\pm}\rangle = 0$. Equations~(\ref{fringes1}) and (\ref{fringes2}) also show that the interference---and hence the anticorrelation---disappears when the pulses no longer overlap in time at the second 50-50 beam splitter, because 
\begin{equation}
|\delta t| >> \tau_p \Rightarrow  f^*(t_+)  f(t_-)=0,\,\,  \forall\, t. 
\end{equation}

At this point we have accomplished our objective.  Our simple classical-field theory predicts singles rates that are independent of the differential delay, and a coincidence rate that exhibits a pronounced dip (anticorrelation) within a (post interference-filter) laser pulse duration, in agreement with the experimental results from \cite{Chen}.  We shall close by delving a little deeper into how our work stacks up against those experiments.  We have assumed $N\ll 1$, i.e., that the average photon number coupled into each fiber from a single laser pulse is much smaller than one. Our theory gives
\begin{equation}
\max(C_{12})/S_1 = 3\eta N/2.
\end{equation}
From Fig.~4 of \cite{Chen} we then get $3\eta N/2 \approx 0.004$ that, for reasonable values of $\eta$ (say, $\eta \sim 0.1$), is consistent with $N$ being much smaller than one \cite{footnote5}.

Our theory predicts that the anticorrelation dip has visibility
 \begin{equation}
 \mathcal{V} \equiv \frac{\max(C_{12})-\min(C_{12})}{ \max(C_{12}) + \min(C_{12})} = 1/5,
 \end{equation}
which is in reasonable agreement with the experimental results from Fig.~4 of \cite{Chen}.   If we eliminate accidental coincidences (the terms that Chen \em et al\/\rm.\ refer to as ``self-intensity correlations'') from our theory---by subtracting from $C_{12}$ in Eq.~(\ref{Cab}) the coincidence rate when $E_+(t)=0$ and the coincidence rate when $E_-(t) = 0$---we get
\begin{equation}
C_{12} \approx \frac{\eta^2N^2}{2}\left(1-e^{-\delta t^2/\tau_p^2}\right),
\end{equation}
which implies the anticorrelation dip has perfect, $\mathcal{V} = 1$, visibility.  When Chen \em et al\/\rm.\ do the like correction to their anticorrelation data, they find near-unity visibility, in agreement with our theory.

In conclusion, we have provided a classical explanation for the anticorrelation experimental results reported in \cite{Chen}.  Thus those experimental results do \em not\/\rm\ require us to abandon the well-accepted precepts that laser light through linear-optics transformations can be modeled as a coherent-state or a classical mixture of coherent states, and that the photodetection measurement statistics for such states can be computed from semiclassical theory, in which the light is treated classically.

The work of J.H.S. was supported by the DARPA Information in a Photon Program under U.S. Army Research Office Grant No.\ W911NF-10-1-0404.


\begin{thebibliography}{2}

\bibitem{Chen}H. Chen, T. Peng, S. Karmakar, Z. Xie, and Y. Shih, Phys. Rev. A {\bf 84,} 033835 (2011).
\bibitem{footnote1}In the semiclassical theory of photodetection, light is taken to be a classical electromagnetic wave, and the discreteness of the electron charge  leads to shot noise as the fundamental noise in photodetection.  Because the shot noises from physically independent photodetectors are statistically independent, the explanation we will present below for the anticorrelation seen in \cite{Chen} will derive solely from randomness in classical electromagnetic fields.
\bibitem{Shapiro}J. H. Shapiro, IEEE J. Sel. Top. Quantum Electron. {\bf 15,}1547 (2009).
\bibitem{MandelWolf}L. Mandel and E. Wolf, \em Optical Coherence and Quantum Optics\/\rm\ (Cambridge Univ. Press, Cambridge, 1995).
\bibitem{footnote2}This distance is also the characteristic size of the speckles cast by the diffuser, see J. W. Goodman, \em Speckle Phenomena in Optics:  Theory and Applications\/\rm\
(Roberts \& Co., Englewood Colo. 2007).
\bibitem{footnote3}Throughout our analysis we will neglect all propagation delays with the exception of the interferometer's differential delay, $\delta t$.
\bibitem{footnote4}This same result can be obtained from the Gaussian moment-factoring theorem, which implies that $\langle | E_1(t)|^2| E_2(u)|^2\rangle$ = $\langle | E_1(t)|^2\rangle\langle | E_2(u)|^2\rangle$ + $|\langle E^*_1(t) E_2(u)\rangle|^2$ for the case at hand.  It is here where the classical-field analysis in Chen \em et al\/\rm. goes astray.  In particular,  our Eqs.~(\ref{Epm}), (\ref{EaDefn}), and (\ref{EbDefn}) yield $\langle E^*_1(t) E_2(u)\rangle = N[f^*(t_+)f(u_+) - f^*(t_-)f(u_-)]/2$, which is dependent on the interferometer's differential delay $\delta t$.  In contrast,  Chen \em et al\/\rm. incorrectly assert, below their Eqs.~(28) and (29), that no such $\delta t$ dependence exists in the corresponding terms from their moment-factoring analysis.  [The reader should note, in this regard, that our $E_+(t)$ and $E_-(t)$ correspond, respectively, to $ E_A(t)$ and $ E_B(t)$ from Chen \em et al\/\rm.]
\bibitem{WJ}J. M. Wozencraft and I. M. Jacobs, \em Principles of Communication Engineering\/\rm\ (Wiley, New York 1965).
\bibitem{footnote5}Strictly speaking, our classical-field theory only needs $\eta N\ll 1$, rather than $N\ll 1$, so that no supposition about $\eta$ is actually necessary.

\end{thebibliography}
\end{document}